\documentclass{emulateapj}

\newcommand{\ped}[2]{#1_\mathrm{#2}}
\newcommand{\p}{\phantom{-}}

\slugcomment{Accepted for publication to ApJ Letters}

\shorttitle{SN\,2003lw and GRB\,031203}
\shortauthors{Malesani et al.}

\begin{document}

\title{SN\,2003lw and GRB\,031203: A Bright Supernova for a Faint
Gamma-ray Burst}

\author{
  D. Malesani\altaffilmark{1},
  G. Tagliaferri\altaffilmark{2},
  G. Chincarini\altaffilmark{2,3},
  S. Covino\altaffilmark{2},
  M. Della Valle\altaffilmark{4},
  D. Fugazza\altaffilmark{2},
  P.~A. Mazzali\altaffilmark{5,6},
  F.~M. Zerbi\altaffilmark{2},
  P. D'Avanzo\altaffilmark{2},
  S. Kalogerakos\altaffilmark{2,7},
  A. Simoncelli\altaffilmark{2,7},
  L.~A. Antonelli\altaffilmark{8},
  L. Burderi\altaffilmark{8},
  S. Campana\altaffilmark{2},
  A. Cucchiara\altaffilmark{2,7},
  F. Fiore\altaffilmark{8},
  G. Ghirlanda\altaffilmark{2},
  P. Goldoni\altaffilmark{9},
  D. G\"otz\altaffilmark{10,3},
  S. Mereghetti\altaffilmark{10},
  I.~F. Mirabel\altaffilmark{9,11},
  P. Romano\altaffilmark{2},
  L. Stella\altaffilmark{8},
  T. Minezaki\altaffilmark{12},
  Y. Yoshii\altaffilmark{12},
  K. Nomoto\altaffilmark{6}
}
\altaffiltext{1}{International School for Advanced Studies (SISSA-ISAS),
  via Beirut 2-4, I-34014 Trieste, Italy.}
\altaffiltext{2}{INAF --- Osservatorio Astronomico di Brera, via
  E. Bianchi 46, I-23807 Merate (Lc), Italy.}
\altaffiltext{3}{Universit\`a degli studi di Milano-Bicocca,
  Dipartimento di Fisica, Piazza delle Scienze 3, I-20126 Milano,
  Italy.}
\altaffiltext{4}{INAF --- Osservatorio Astronomico di Arcetri, largo
  E. Fermi 5, I-50125 Firenze, Italy.}
\altaffiltext{5}{INAF --- Osservatorio Astronomico di Trieste, via
  G. Tiepolo 11, I-34131 Trieste, Italy.}
\altaffiltext{6}{Department of Astronomy, School of Science, University
  of Tokyo, Bunkyo-ku, Tokyo 113-0033, Japan.}
\altaffiltext{7}{Universit\`a degli Studi di Milano, Dipartimento di
  Fisica, via Celoria 16, I-20133 Milano, Italy.}
\altaffiltext{8}{INAF --- Osservatorio Astronomico di Roma, via di
  Frascati 33, I-00040 Monteporzio Catone (Roma), Italy.}
\altaffiltext{9}{CEA Saclay, DSM/DAPNIA/Service d'Astrophysique,
  F-91191, Gif-sur-Yvett, France.}
\altaffiltext{10}{IASF-CNR, Ist. di Astrofisica Spaziale e Fisica
  Cosmica, Sez. di Milano, via Bassini 15, I-20133 Milano, Italy.}
\altaffiltext{11}{Instituto de Astronomia y Fisica del Espacio, CC 67, Suc 28,
  1428 Capital Federal, Argentina.}
\altaffiltext{12}{Institute of Astronomy, School of Science, University
  of Tokyo, 2-21-1 Osawa, Mitaka, Tokyo 181-0015, Japan.}


\begin{abstract}
Optical and near-infrared observations of the gamma-ray burst
GRB\,031203, at $z = 0.1055$, are reported. A very faint afterglow is
detected superimposed to the host galaxy in our first infrared $JHK$
observations, carried out $\sim 9$~hours after the burst. Subsequently,
a rebrightening is detected in all bands, peaking in the $R$ band about
18 rest-frame days after the burst. The rebrightening closely resembles
the light curve of a supernova like SN\,1998bw, assuming that the GRB
and the SN went off almost simultaneously, but with a somewhat slower
evolution. Spectra taken close to the maximum of the rebrightening show
extremely broad features as in SN\,1998bw. The determination of the
absolute magnitude of this SN (SN\,2003lw) is difficult owing to the
large and uncertain extinction, but likely this event was brighter than
SN\,1998bw by $0.5$~mag in the $VRI$ bands, reaching an absolute
magnitude $M_V = -19.75 \pm 0.15$.
\end{abstract}



\section{Introduction}

In recent years, extensive optical and near-infrared (NIR) follow-up of
gamma-ray bursts (GRBs) has revealed a physical connection between a
significant fraction of long-duration GRBs and core-collapse
supernov\ae{} (SNe). First, the bright SN\,1998bw was discovered
spatially and temporally coincident with GRB\,980425
\citep{Ga98,Ku98}. However, GRB\,980425 was rather different from
classical, cosmological GRBs, being severely underenergetic and lacking
an optical afterglow. Then, SN\,2003dh was detected in the afterglow of
GRB\,030329 \citep{St03,Hj03}. Both SNe showed broad bumps in their
spectra, indicating very large expansion velocities (up to
$30\,000$~km/s), and were extremely bright.  These highly-energetic SNe
are often named hypernov\ae{} \citep[e.g.][]{Iw98}. Last, bumps
discovered in the light curves of several afterglows, peaking $\sim
20$~days after the GRB, have been interpreted as due to the emerging of
SNe out of the afterglow light, based on their brightness, temporal
evolution and colors \citep[e.g.][]{Bl99,Ga03}. The bumps resemble the
light curve of SN\,1998bw, with a certain scatter in the brightness and
rise time \citep[e.g.][]{Ze04}. Spectroscopic confirmation that the bump
of GRB\,021211 has a SN spectrum \citep[SN\,2002lt;][]{DV03} supports
this conclusion. These observations indicate that the GRB/SN association
is common.


GRB\,031203 was discovered by the INTEGRAL satellite on 2003 Dec 3.91769
UT \citep{Go03}, with a duration of $\sim 30$~s and a peak flux of $1.3
\times 10^{-7}$~erg~cm$^{-2}$~s$^{-1}$
\citep[$20-200$~keV;][]{Me03a}. The precise and fast dissemination of
the GRB coordinates by the INTEGRAL burst alert system \citep{Me03b}
allowed an effective search for the afterglow. We also immediately
activated our ToO program at ESO, starting NIR observations at the NTT
7~hours after the GRB \citep{Ze03}. The X-ray and radio afterglows were
soon discovered \citep{Sa03,Fr03}. A compact galaxy, located at a
consistent position,
was proposed to be the GRB host galaxy by \citet{Pr03}. The redshift was
$z = 0.1055 \pm 0.0001$ \citep{Pr03,Pr04}, making GRB\,031203 the second
closest burst after GRB\,980425 at $z = 0.0085$ \citep{Ga98}.
\citet{Va04} discovered a scattered expanding X-ray halo due to the
reflection of the burst and/or early afterglow light from Galactic dust
grains. This allowed an (indirect) measurement of the X-ray flux at the
earliest stages after the burst onset.

Given the low redshift of this event, the isotropic-equivalent burst
energy is extremely low%
\footnote{We adopt a cosmology with $H_0 = 71$~km~s$^{-1}$~Mpc$^{-1}$,
$\ped{\Omega}{m} = 0.27$, $\Omega_\Lambda = 0.73$ (WMAP results). At $z
= 0.1055$ the luminosity distance is $D =477$~Mpc and the distance
modulus is $\mu = 38.42$~mag.},
$\ped{E}{iso} \sim 3 \times 10^{49}$~erg
\citep[20-2000~keV;][]{Wa04,Pr04}, well below the standard reservoir 
$\sim 2 \times 10^{51}$~erg of normal GRBs \citep{Fr01,Bl03}. Only
GRB\,980425 \citep{Ga98} and XRF\,020903 \citep{Sa04} were less
energetic.

Based on photometric monitoring of the host galaxy, several groups have
reported evidence for a SN associated with GRB\,031203
\citep{Be04,Th04,Co04,Ga04}. After the ultimate confirmation, coming from
spectroscopic observations and reported by our group \citep{Ta04}, the
IAU named this event SN\,2003lw.

\section{Observations and data reduction}

{\bf Photometry.} We observed the field of GRB\,031203 starting $\approx
7$~h after the trigger, to search for the near-infrared (NIR) afterglow,
using SofI on the ESO-NTT at La Silla (Chile). Subsequent imaging with
ISAAC on the ESO-VLT showed the presence of a varying source coincident
with the putative host galaxy of GRB\,031203: the total flux had dimmed
in the $J$, $H$ and $K$ filters by a few tenths of a magnitude (see
Fig.~\ref{fg:lc}). We therefore started a campaign to monitor the
optical/NIR light curve of the event, searching for a SN
rebrightening. The observing log is presented in Tab.~\ref{tb:phot}.

Image reduction and analysis were performed following the standard
procedures, by means of both aperture photometry and PSF-matched image
subtraction.  To avoid saturation from a nearby bright star, the
exposure time was always kept short. In same cases, occulting bars were
placed to cover the bright star (showing that the effect was
negligible). Optical and NIR photometry were calibrated against Landolt
standard stars and the 2MASS, respectively. To focus on the issue of
variability, in Tab.~\ref{tb:phot} we list just the relative photometry
with respect to a reference epoch. We should also note that the host
galaxy spectrum is dominated by prominent emission lines. This may lead
to relatively large offsets when comparing results from other
instruments, owing to unavoidable small differences in the filter
profiles and CCD efficiencies.

Additional $K$ and $I$ photometry was acquired with the 2~m MAGNUM
telescope of the University of Tokyo \citep{Yo03}, located in the Hawaii
Islands. Although the different shape of the MAGNUM and ESO $I$ filters
(particularly critical due to the presence of the bright H$\alpha$ line
in the blue filter wing) makes it difficult to compare the results, the
data are in good agreement (Tab.~\ref{tb:phot}). For consistency, these
data are not plotted in Fig.~\ref{fg:lc}.

\begin{deluxetable}{llllll}
\tablewidth{\columnwidth}
\tabletypesize{\scriptsize}
\tablecaption{Summary of photometric observations.\label{tb:phot}}
\tablehead{
  \colhead{UT start\tablenotemark{a}} &
  \colhead{Seeing} & \colhead{Instrument} & \colhead{Band} &
  \colhead{Magnitude\tablenotemark{b}}}%
\startdata
031220.247 & 0.3$\arcsec$ & FORS2 & $V$ &  -0.023$\pm$0.020 \\
\textbf{031230.250} & 0.5$\arcsec$ & FORS1 & $V$ & \p\textbf{20.37$\pm$0.05} \\
040228.193 & 0.7$\arcsec$ & FORS1 & $V$ & \p0.165$\pm$0.012 \\
040302.064 & 0.7$\arcsec$ & FORS1 & $V$ & \p0.169$\pm$0.015 \\
\textbf{031215.314} & 0.6$\arcsec$ & FORS1 & $R$ & \p\textbf{20.14$\pm$0.03} \\
031217.284 & 0.7$\arcsec$ & FORS1 & $R$ &  -0.009$\pm$0.016 \\
031223.295 & 0.5$\arcsec$ & FORS1 & $R$ &  -0.060$\pm$0.016 \\
031228.296 & 0.6$\arcsec$ & FORS1 & $R$ &  -0.051$\pm$0.010 \\
031230.241 & 0.5$\arcsec$ & FORS1 & $R$ &  -0.044$\pm$0.013 \\
040116.171 & 0.8$\arcsec$ & FORS1 & $R$ & \p0.130$\pm$0.013 \\
040117.185 & 0.5$\arcsec$ & FORS1 & $R$ & \p0.139$\pm$0.014 \\
040221.101 & 0.7$\arcsec$ & FORS1 & $R$ & \p0.248$\pm$0.019 \\
040228.198 & 0.7$\arcsec$ & FORS1 & $R$ & \p0.242$\pm$0.012 \\
040302.069 & 0.6$\arcsec$ & FORS1 & $R$ & \p0.255$\pm$0.016 \\
\textbf{031215.321} & 0.5$\arcsec$ & FORS1 & $I$ & \p\textbf{19.18$\pm$0.03} \\
031217.291 & 0.6$\arcsec$ & FORS1 & $I$ &  -0.014$\pm$0.015 \\
031223.300 & 0.4$\arcsec$ & FORS1 & $I$ &  -0.052$\pm$0.013 \\
031228.301 & 0.6$\arcsec$ & FORS1 & $I$ &  -0.044$\pm$0.015 \\
031230.245 & 0.4$\arcsec$ & FORS1 & $I$ &  -0.052$\pm$0.009 \\
040116.177 & 0.6$\arcsec$ & FORS1 & $I$ & \p0.050$\pm$0.008 \\
040117.179 & 0.6$\arcsec$ & FORS1 & $I$ & \p0.054$\pm$0.007 \\
040221.107 & 0.5$\arcsec$ & FORS1 & $I$ & \p0.151$\pm$0.013 \\
040228.203 & 0.8$\arcsec$ & FORS1 & $I$ & \p0.150$\pm$0.013 \\
\textbf{031204.288} & 0.9$\arcsec$ & SofI  & $J$ & \p1\textbf{8.13$\pm$0.034} \\
031205.258 & 0.5$\arcsec$ & ISAAC & $J$ & \p0.143$\pm$0.030 \\
031214.140 & 1.1$\arcsec$ & SofI  & $J$ & \p0.014$\pm$0.032 \\
031223.191 & 1.0$\arcsec$ & SofI  & $J$ &  -0.084$\pm$0.069 \\
040228.111 & 0.5$\arcsec$ & ISAAC & $J$ & \p0.110$\pm$0.045 \\
\textbf{031204.300} & 0.9$\arcsec$ & SofI  & $H$ & \p\textbf{17.36$\pm$0.042} \\
031205.271 & 0.5$\arcsec$ & ISAAC & $H$ & \p0.257$\pm$0.035 \\
031214.148 & 0.9$\arcsec$ & SofI  & $H$ &  -0.006$\pm$0.033 \\
040228.104 & 0.5$\arcsec$ & ISAAC & $H$ & \p0.190$\pm$0.019 \\
\textbf{031204.204} & 0.9$\arcsec$ & SofI  & $K$ & \p\textbf{16.38$\pm$0.036} \\
031204.312 & 0.8$\arcsec$ & SofI  & $K$ &  -0.036$\pm$0.033 \\
031205.267 & 0.5$\arcsec$ & ISAAC & $K$ & \p0.161$\pm$0.025 \\
031214.154 & 0.8$\arcsec$ & SofI  & $K$ & \p0.046$\pm$0.044 \\
031223.188 & 1.0$\arcsec$ & SofI  & $K$ & \p0.082$\pm$0.107 \\
040228.095 & 0.5$\arcsec$ & ISAAC & $K$ & \p0.255$\pm$0.055 \\[\smallskipamount]
031224.507 & 1.5$\arcsec$ &MAGNUM & $I$ & \p19.16$\pm$0.03 \\
040106.447 & 1.4$\arcsec$ &MAGNUM & $I$ & \p19.16$\pm$0.03 \\
031224.503 & 1.1$\arcsec$ &MAGNUM & $K$ & \p16.61$\pm$0.04 \\
040106.445 & 1.1$\arcsec$ &MAGNUM & $K$ & \p16.58$\pm$0.04 \\
\enddata

\tablenotetext{a}{In the form \textit{yymmdd}, for year, month, day.}
\tablenotetext{b}{Magnitudes are given relative to the boldface epoch.}

\end{deluxetable}

{\bf Spectroscopy.} Moderate-resolution spectra (FWHM $\mbox{}\approx
10$~\AA) were taken with the VLT on 2003 Dec.~20 (FORS\,2), 2003 Dec.~30
(FORS\,1), and 2004 Mar.~01 (FORS\,1).  Flux calibration was achieved by
observing spectrophotometric stars.  After comparing synthetic
magnitudes calculated from our spectra with the photometry we introduced
a correction to account for light loss outside the slit. To ensure a
sound relative calibration between the spectra, we also checked that the
fluxes of the host galaxy emission lines did not vary. A detailed
discussion of the spectroscopy and of the host galaxy will be presented
elsewhere (Chincarini et al. 2004, in preparation; hereafter C04).

\section{Results and discussion}

\begin{figure}\centering
  \includegraphics[width=\columnwidth]{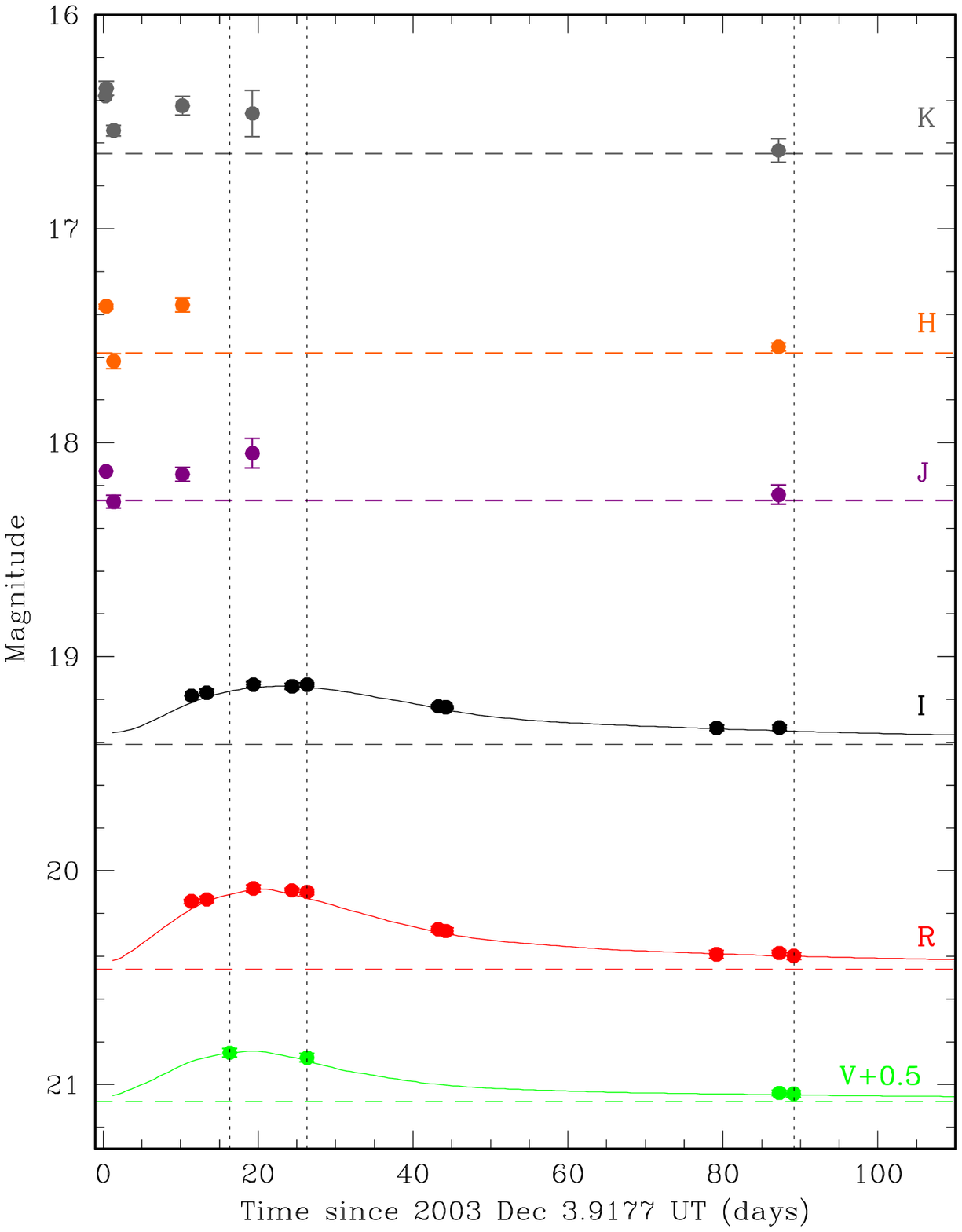}
  \caption{Optical and NIR light curves of GRB\,031203 (dots). Error
  bars indicate the amount of relative errors only
  (Tab.~\ref{tb:phot}). The solid curves show the evolution of
  SN\,1998bw \citep{Ga98,Mc00}, rescaled at $z = 0.1055$, stretched by a
  factor 1.1, extinguished with $E_{B-V} = 1.1$ and brightened by
  0.5~mag. Dashed lines indicate the host galaxy contribution. Vertical
  lines mark the epochs of our spectra.\label{fg:lc}}
\end{figure}

\begin{figure}\centering
  \includegraphics[width=\columnwidth,keepaspectratio]{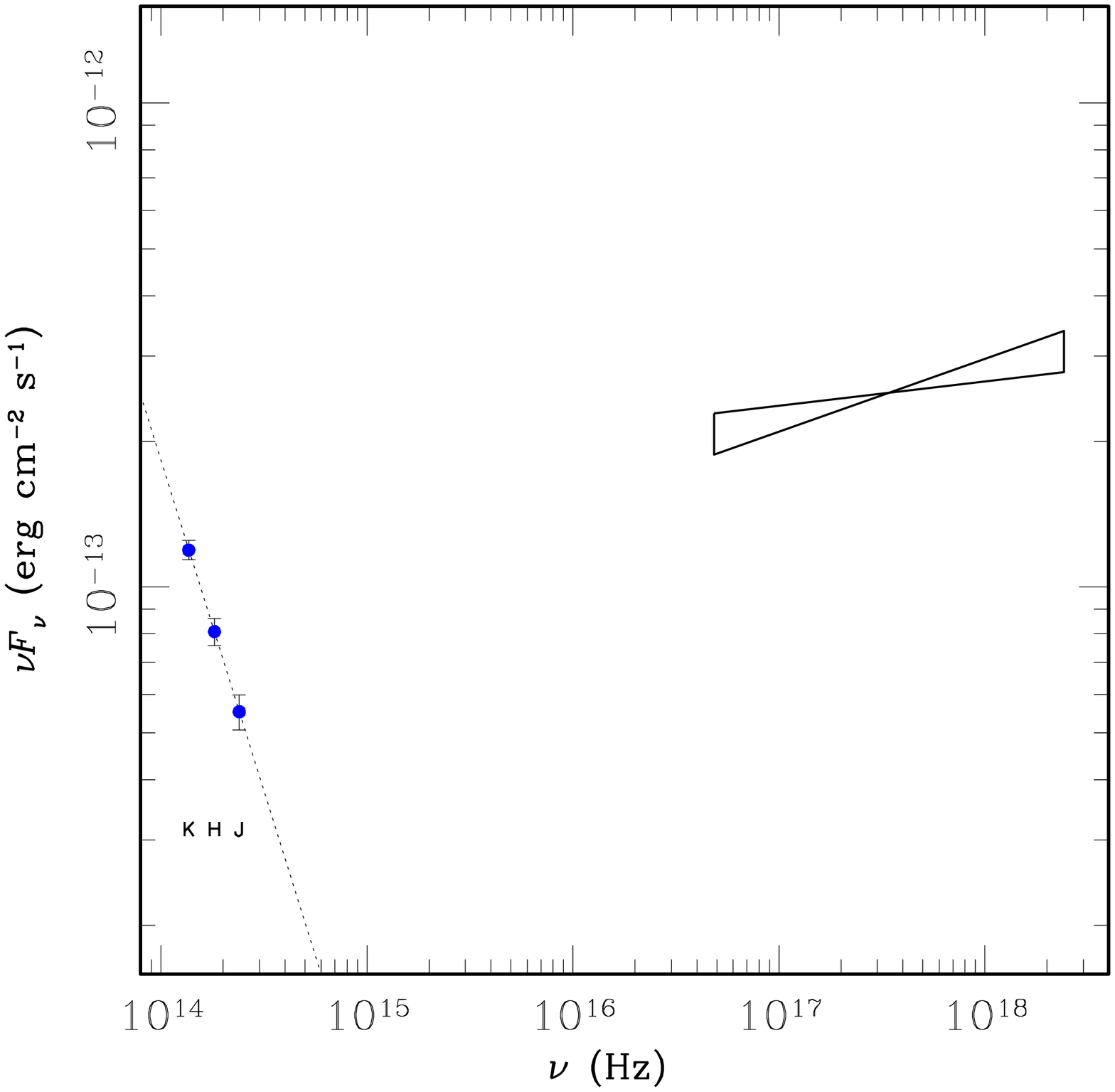}
  \caption{Spectral energy distribution of the afterglow of GRB\,031203
  on 2003 Dec.~4.3 UT (9 hours after the trigger). The NIR values are
  calculated from our data by subtracting the host contribution and
  assuming $E_{B-V} = 1.1$. The NIR spectral index is $\beta = 2.36 \pm
  0.02$ ($F_\nu \propto \nu^{-\beta}$). The X-ray spectrum is from
  \citet{Wa04} (reported at $t = 9$~h using the X-ray decay slope), who
  find $\ped{N}{H} = 8.8 \times 10^{21}$~cm$^{-2}$ and $\ped{\beta}{X} =
  0.90 \pm 0.05$.\label{fg:SED}}
\end{figure}

In Fig.~\ref{fg:lc} we show the light curves of GRB\,031203. Early-time
NIR photometry shows a dimming in all bands between the first and second
night after the GRB. This is confirmed by PSF-matched image
subtraction. We believe that we have seen the NIR afterglow of
GRB\,031203. The magnitudes are $J = 20.60 \pm 0.09$, $H = 19.05 \pm
0.07$, $K = 17.56 \pm 0.05$ (9~hours after the GRB), obtained by
subtracting the host contribution. \citet{Co04} have $I$-band
observations at similar epochs, and do not report evidence for
variability. However, extrapolation to the visible region yields $I \sim
23.4$, quite a faint value when compared to the host luminosity $I
\approx 19.4$. Little contribution from the afterglow is seen in our
measurement of Dec.~5, implying a quick decay between the two nights
($F(t) \propto t^{-\alpha}$, with $\alpha \gtrsim 2$).  However, there
is no variation between the two $K$-band observations of the first night
(separated by 2.6~h), suggesting a break in the light curve or a bumpy
behaviour. In Fig.~\ref{fg:SED}, we compare the spectrum in the NIR and
X-ray regions \citep{Wa04}. A discontinuity is apparent, indicating a
different origin for the emission in the two bands. The X-ray component
has a much harder spectrum, and a slower decay ($\alpha = 0.55 \pm
0.05$). Interestingly, \citet{Wa04} infer a fast decay of the early-time
X-ray afterglow, consistent with our NIR value ($\alpha \gtrsim
1.7$). In the standard model of afterglows \citep[e.g.][]{SPN98}, a fast
decay is consistent with a soft spectrum bluewards of the peak
frequency.

A few days after the GRB, a rebrightening is apparent in all optical/NIR
bands. The rebrightening amounts to $\approx 30\%$ of the total flux,
and is coincident with the center of the host galaxy to within
$0.1\arcsec$ ($\approx 200$~pc at $z = 0.1055$). For comparison, we show
in Fig.~\ref{fg:lc} the $VRI$ light curves of SN\,1998bw
\citep{Ga98,Mc00}, placed at $z = 0.1055$ and dereddened with $E_{B-V} =
1.1$ (see below). Interpolation of the $UBVRI$ data was performed in
order to estimate the fluxes of SN\,1998bw at the frequencies
corresponding to the observed bands. Even after correcting for
cosmological time dilation, the light curve of SN\,2003lw is broader
than that of SN\,1998bw, and requires an additional stretching factor of
$\approx 0.9$ to match the $R$ and $I$ bands. Near the peak, the light
curve is rather flat, resembling the hypernova SN\,1997ef \citep{Iw00}
more than SN\,1998bw. The $R$-band maximum is reached on approximately
2003 Dec.~24 ($\sim 18$ comoving days after the GRB). We note that
the details of the light curve shape are sensitive to the removal of the
host contribution. This may explain the different finding of
\citet{Th04}, who need no stretch, and \citet{Co04}, who find a longer
rise. Assuming a light curve shape similar to SN\,1998bw, which had a
rise time of 16~days in the $V$ band, our data suggest an explosion time
nearly simultaneous with the GRB. However, given that SN\,2003lw was not
strictly identical to SN\,1998bw, and as we lack optical data in the
days immediately following the GRB, a lag of a few days cannot be ruled
out. Type-Ic SNe usually reach $V$-band maximum in $\sim 12$-20 days,
the brightest events showing a slower evolution \citep[see e.g. Fig.~2
of][]{Maz02}.

A precise determination of the absolute magnitude of the SN is made
difficult by the uncertain, and significant, extinction.  C04 and
\citet{Pr04} constrain the average combined Galactic and host extinction
to be $E_{B-V} \approx 1.1$ based on the Balmer ratios of the host
galaxy. Given the good spatial coincidence of the SN with the center of
the host, such value is likely a good estimate for the SN extinction. We
also adopt a Galactic extinction law \citep{Ca89} with $R_V = 3.1$.
With the assumed reddening, SN\,2003lw appears brighter than SN\,1998bw
by 0.5~mag in the $V$, $R$, and $I$ bands. The absolute magnitudes of
SN\,2003lw are hence $M_V = -19.75\pm0.15$, $M_R = -19.9\pm0.08$, and
$M_I = -19.80\pm0.12$. \citet{Th04}, using $I$-band data, also found
that SN\,2003lw was brighter than SN\,1998bw by $\sim 0.55$~mag, in full
agreement with our result. \citet{Co04}, however, found a comparable
luminosity for the two SNe; this discrepancy is entirely due to the
lower extinction they assume.

\begin{figure}\centering
  \includegraphics[width=\columnwidth,keepaspectratio]{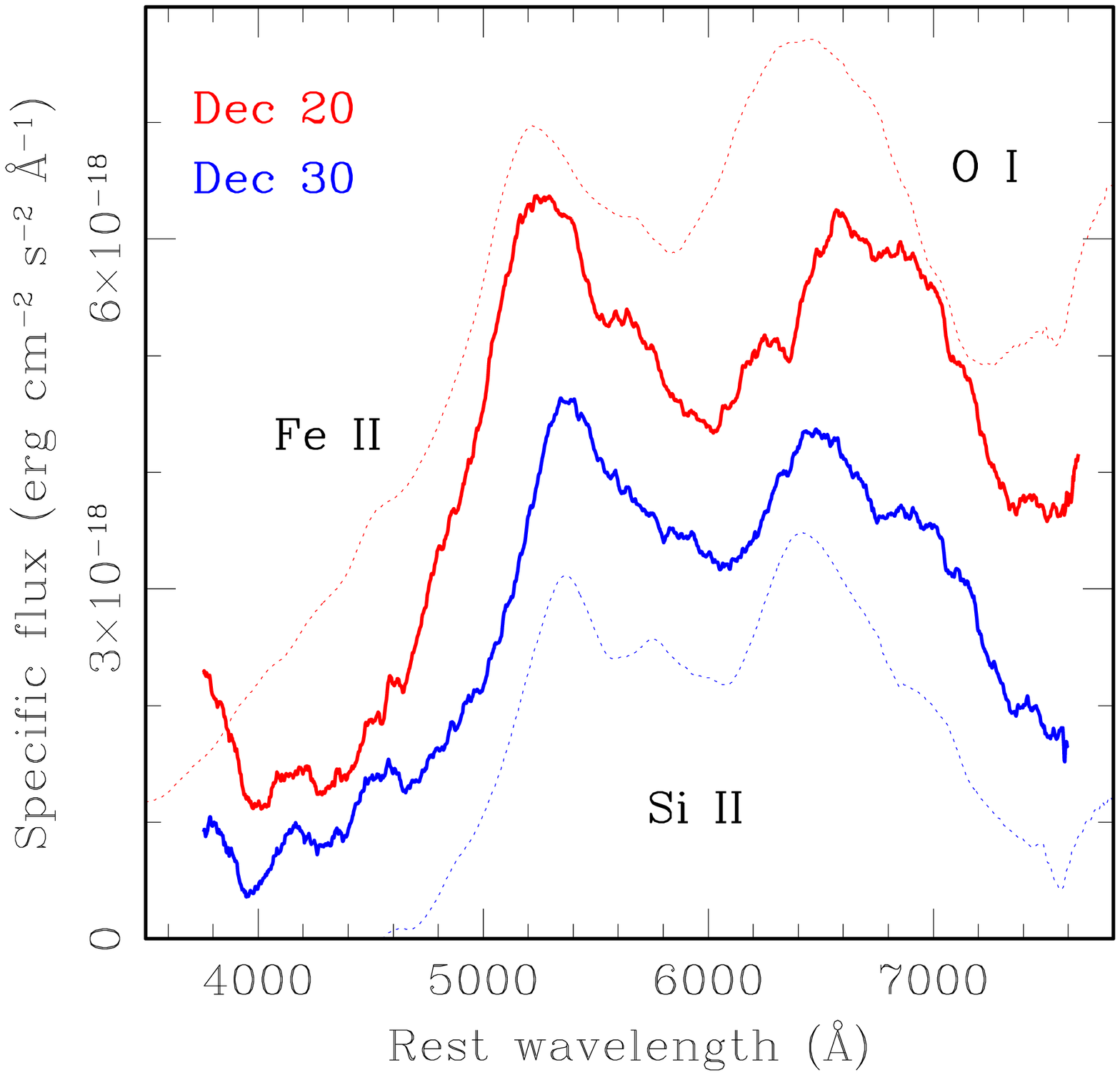}
  \caption{Spectra of SN\,2003lw, taken on 2003 Dec.~20 and Dec.~30
  (solid lines), smoothed with a boxcar filter 250~\AA{} wide.  Dotted
  lines show the spectra of SN\,1998bw \citep[from][]{Pa01}, taken on
  1998 May~9 and May~19 (13.5 and 23.5~days after the GRB, or 2 days
  before and 7~days after the $V$-band maximum), extinguished with
  $E_{B-V} = 1.1$ and a Galactic exinction law \citep{Ca89}. The spectra
  of SN\,1998bw were vertically displaced for presentation
  purpose.\label{fg:spec}}
\end{figure}

Fig.~\ref{fg:spec} shows the spectra of the rebrightening on 2003
Dec.~20 and Dec.~30 (14 and 23 rest-frame days after the GRB), after
subtracting the spectrum taken on 2004 Mar.~1 (81 rest-frame days after
the GRB). This assumes that the latter spectrum contains only a
negligible contribution from the SN, which is confirmed by the
photometry (Fig.~\ref{fg:lc}). The spectra of SN\,2003lw are remarkably
similar to those of SN\,1998bw obtained at comparable epochs
\citep[shown as dotted lines in Fig.~\ref{fg:spec}; from][]{Pa01}. Both
SNe show very broad absorption features, indicating large expansion
velocities. Thus we tentatively classify SN\,2003lw as a hypernova. The
main absorptions are identified in Fig.~\ref{fg:spec} as in SN\,1998bw,
following \citet{Iw98}. The velocity of the Si\,II line in SN\,2003lw is
apparently smaller than in SN\,1998bw. The broad peaks near 5300~\AA{}
and 6600~\AA{} are probably the emission components of P-Cygni profiles
due to the blending of several lines. There is evolution between the two
epochs: the bluer bump is observed at longer wavelengths in the second
spectrum, and is slighty narrower. Moreover, the shape of the redder
peak is different in the two epochs.  Both peaks appear at redder
wavelengths than in SN\,1998bw. Detailed modeling of the spectra will be
presented elsewhere (Mazzali et al. 2004, in preparation).\bigskip

By modeling the X-ray dust echo, \citet{Wa04} concluded that GRB\,031203
was an X-ray flash (XRF); however, the prompt emission data do not
confirm this hypotesis (Sazonov, Lutovinov \& Sunyaev 2004, in
preparation). This event, like SN\,1998bw \citep{Pi00}, seems therefore
to violate the correlation between the isotropic-equivalent gamma-ray
energy $\ped{E}{iso}$ and the peak spectral energy $\ped{E}{p}$
\citep{Am02,La03}. In fact, assuming $\ped{E}{iso} \sim 1.5
\times 10^{50}$~erg \citep[1-10000~keV;][]{Wa04}, the \citet{Am02}
relation would imply $\ped{E}{p} \sim 10$~keV, a value indicating an XRF
nature for GRB\,031203. This is in contrast with INTEGRAL data. Of
course, this issue will be be settled only after a thorough analysis of
the prompt emission spectra.

The afterglow of GRB\,031203 was very weak, the faintest ever detected
in the optical/NIR. Extrapolation in the $R$ band yields a luminosity
$\sim 200$~times fainter than the dimmest afterglow discovered so far
\citep[GRB\,021211:][]{Fo03,Pa03}. The detection of the SN optical light
implies that the reason of such faintness was not an extreme dust
obscuration. Also given the low redshift of the event, this example
shows that some optical afterglows may escape detection just because
they are faint \citep[e.g.][]{Fy01,La02,DP03}.

GRB\,031203, together with GRB\,980425 at $z = 0.085$, was a very dim
event, perhaps a jet observed far from its axis
\citep[e.g.][]{Ma02,Ya03}. Being so faint, they would have been likely
missed at cosmological distances. Since the volume they sample is much
smaller than that probed by classical, distant GRBs with $\langle z
\rangle \approx 1$, the rate of these events could be much larger.
As noted by \citet{Th04}, this would increase the detection rate for the
{\it Swift} satellite \citep{Ge04}. More rapid and efficient
observations, also soon feasible thanks to {\it Swift}, will allow a
detailed study of this largely unexplored class of events.

GRB\,031203 was quite similar to GRB\,980425, even if overall more
powerful. Both events consisted in a single, underenergetic pulse. Their
afterglows were very faint or absent in the optical, and showed a very
slow decline in the X-ray \citep{Pi00,Wa04}. Last, they were both
accompained by a powerful hypernova.


\acknowledgments

We thank the anonymous referee for useful suggestions and prompt
reply. The data presented in this work were obtained with ESO telescopes
under programmes 072.D-0480 and 072.D-0137. We are grateful to the ESO
staff, and in particular to M.\,Billeres, O.\,Hainaut, S.\,Hubrig,
R.\,Johnson, C.\,Lidman, G.\,Marconi, and T.\,Szeifert.  F.\,Patat is
warmly acknowledged for providing the SN\,1998bw data. DM thanks
L.\,Ballo for useful discussions. This work was supported by contract
ASI/I/R/390/02 of the Italian Space Agency for the Italian {\it Swift}
Project, by the Italian Ministry for University and Research and by the
Italian National Institute for Astrophysics (INAF).


\begin{figure}\centering
  \includegraphics[width=\textwidth]{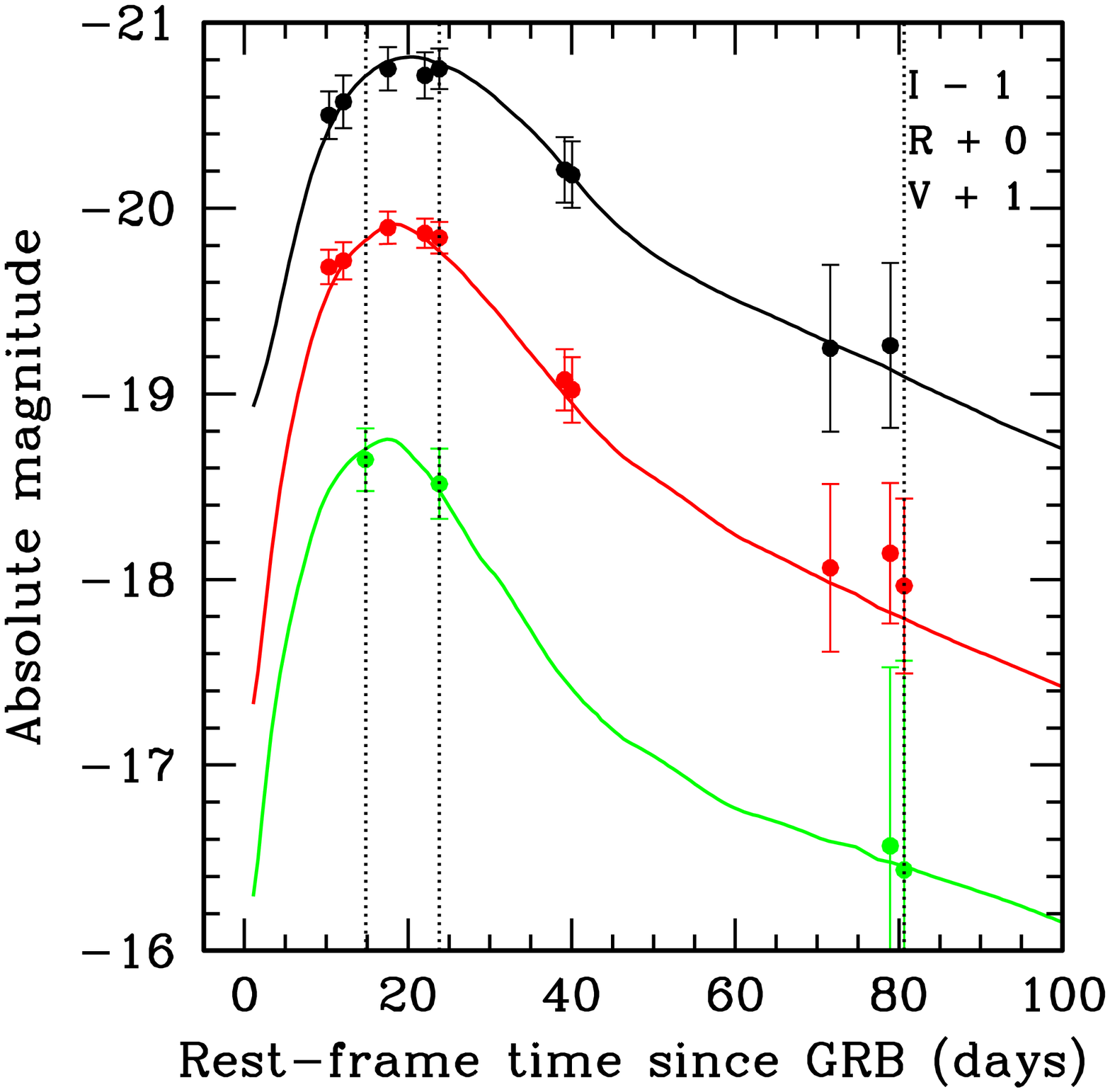}
  \caption{Light curves of SN\,2003lw in the observed $V$, $R$, and $I$
  filters, after removing the host galaxy contribution. Solid lines show
  the light curves of SN\,1998bw \citep{Ga98,Mc00} reported at the
  redshift of SN\,2003lw, extinguished with $E_{B-V} = 1.1$, brightened
  by $0.5$~mag, and stretched by a factor 1.1. The host contribution was
  assumed to be $V = 20.57$, $R = 20.47$, $I = 19.41$. NB: this figure
  will not appear on the ApJL version, due to space limitations.}
\end{figure}


\begin{thebibliography}{}

\bibitem[Amati et al.(2002)]{Am02} Amati, L., et al. 2002, \aap, 390, 81
\bibitem[Bersier et al.(2004)]{Be04} Bersier, D., et al. 2004, GCN Circ
  2544
\bibitem[Bloom et al.(1999)]{Bl99} Bloom, J. S., et al. 1998, \nat, 401,
  453
\bibitem[Bloom, Frail \& Kulkarni(2003)]{Bl03} Bloom, J. S., Frail,
  D. A., \& Kulkarni, S. R. 2003, \apj, 594, 674
\bibitem[Cardelli, Clayton \& Mathis(1989)]{Ca89} Cardelli, J. A.,
  Clayton, G. C., \& Mathis, J. S. 1989, \apj, 345, 245
\bibitem[Cobb et al.(2004)]{Co04} Cobb, B.E., Baylin, C.D., van Dokkum,
  P.G., Buxton, M.M., \& Bloom, J.S. 2004, \apj, 608, L93
\bibitem[Della Valle et al.(2003)]{DV03} Della Valle, M., et al. 2003,
  \aap, 406, L33
\bibitem[De Pasquale et al.(2003)]{DP03} De Pasquale, M., et al. 2003,
  \apj, 592, 1018
\bibitem[Fox et al.(2003)]{Fo03} Fox, D. W., et al. 2003, \apj, 586, L5
\bibitem[Frail et al.(2001)]{Fr01} Frail, D. A., et al. 2001, \apj, 562,
  L55
\bibitem[Frail(2003)]{Fr03} Frail, D. A. 2003, GCN Circ 2473
\bibitem[Fynbo et al.(2001)]{Fy01} Fynbo, J. U., et al. 2001, \aap, 369,
  373
\bibitem[Galama et al.(1998)]{Ga98} Galama, T. J., et al. 1998, \nat,
  395, 670
\bibitem[Gal-Yam et al.(2004)]{Ga04} Gal-Yam, A., et al. 2004, \apjl,
  submitted (astro-ph/0403608)
\bibitem[Garnavich et al.(2003)]{Ga03} Garnavich, P. M., et al. 2003,
  \apj, 582, 924
\bibitem[Gehrels et al.(2004)]{Ge04} Gehrels, N., et al. 2004, \apj,
  in press (astro-ph/0405233)
\bibitem[G\"otz et al.(2003)]{Go03} G\"otz, D., Mereghetti, S., Beck,
  M., Borkowski, J., \& Mowlavi, N. 2003, GCN Circ 2459
\bibitem[Heise et al.(2003)]{He03} Heise, J. 2003, in AIP
  Conf. Ser. 662, Gamma-ray Burst and Afterglow Astronomy 2001,
  ed. G. R. Ricker \& R. K. Vanderspek (New York: AIP), 229
\bibitem[Hjorth et al.(2003)]{Hj03} Hjorth, J., et al. 2003, \nat, 423,
  847
\bibitem[Iwamoto et al.(1998)]{Iw98} Iwamoto, K., et al. 1998, \nat,
  395, 672
\bibitem[Iwamoto et al.(2000)]{Iw00} Iwamoto, K., et al. 2000, \apj,
  534, 660
\bibitem[Kulkarni et al.(1998)]{Ku98} Kulkarni, S. R., et al. 1998,
  \nat, 395, 663
\bibitem[Lamb et al.(2003)]{La03} Lamb, D. Q., Donaghy, T. Q., \&
  Graziani, C. 2003, \apj, submitted (astro-ph/0312634)
\bibitem[Lazzati et al.(2002)]{La02} Lazzati, D., Covino, S., \&
  Ghisellini, G. 2002, \mnras, 330, 583
\bibitem[Maeda et al.(2002)]{Ma02} Maeda, K., Nakamura, T., Nomoto, K.,
  Mazzali, P. A., Patat, F., \& Hachisu, I. 2002, \apj, 565, 405
\bibitem[Mazzali et al.(2002)]{Maz02} Mazzali, P. A., et al. 2002, \apj,
  572, L61
\bibitem[McKenzie \& Schaefer(2000)]{Mc00} McKenzie, E. H., \& Schaefer,
  B. E. 2000, \pasp, 111, 964
\bibitem[Mereghetti \& G\"otz(2003a)]{Me03a} Mereghetti, S., \& G\"otz,
  D. 2003a, GCN Circ 2460
\bibitem[Mereghetti et al.(2003b)]{Me03b} Mereghetti, S., G\"otz, D.,
  Borkowski, J., Walter, R., \& Pedersen, H. 2003b, \aap, 411, L291
\bibitem[Pandey et al.(2003)]{Pa03} Pandey, S. B., et al. 2003, \aap,
  408, L21
\bibitem[Patat et al.(2001)]{Pa01} Patat, F., et al. 2001, \apj, 555,
  900
\bibitem[Pian et al.(2000)]{Pi00} Pian, E., et al. 2000, \apj, 536, 778
\bibitem[Prochaska et al.(2003)]{Pr03} Prochaska, J. X., Chen, H. W.,
  Hurley, K., Bloom, J. S., Graham, J. R., \& Vacca, W. D. 2003, GCN
  Circ 2475
\bibitem[Prochaska et al.(2004)]{Pr04} Prochaska, J. X., et al. 2004,
  \apj, in press (astro-ph/0402085)
\bibitem[Sakamoto et al.(2004)]{Sa04} Sakamoto, T., et al. 2004, \apj,
  602, 875
\bibitem[Santos-Lleo \& Calderon(2003)]{Sa03} Santos-Lleo, M., \&
  Calderon, P. 2003, GCN Circ 2464
\bibitem[Sari, Piran \& Narayan(1998)]{SPN98} Sari, R., Piran, T., \&
  Narayan, R. 1998, \apj, 497, L17
\bibitem[Stanek et al.(2003)]{St03} Stanek, K.Z., et al. 2003, \apj,
  591, L17
\bibitem[Tagliaferri et al.(2004)]{Ta04} Tagliaferri, G., et al. 2004,
  IAU Circ 8308
\bibitem[Thomsen et al.(2004)]{Th04} Thomsen, B., et al. 2004, \aap,
  419, L21
\bibitem[Vaughan et al.(2004)]{Va04} Vaughan, S., et al. 2004, \apj,
  603, L5
\bibitem[Watson et al.(2004)]{Wa04} Watson, D., et al. 2004, \apj, 605,
  L97
\bibitem[Yamazaki, Yonetoku \& Nakamura(2003)]{Ya03} Yamazaki, R.,
  Yonetoku, D., \& Nakamaura, T. 2003, \apj, 594, L79
\bibitem[Yoshii et al.(2003)]{Yo03} Yoshii, Y., et al. 2003, \apj, 592,
  467
\bibitem[Zeh, Klose \& Hartmann(2004)]{Ze04} Zeh, A., Klose, S., \&
  Hartmann, D. H. 2004, ApJ, in press (astro-ph/0311610)
\bibitem[Zerbi et al.(2003)]{Ze03} Zerbi, F. M., et al. 2003, GCN Circ
  2471

\end{thebibliography}
\end{document}